\documentclass[10pt,twocolumn,twoside]{IEEEtran}
\usepackage{times,amsmath,amssymb,dsfont,graphicx,float,enumerate}
\newtheorem{Theorem}{Theorem}
\newtheorem{Prop}{Proposition}
\newtheorem{Cor}{Corollary}
\newtheorem{lemma}{Lemma}

\newtheorem{rem}{Remark}
\ifCLASSINFOpdf
\else
\fi
\IEEEoverridecommandlockouts

\begin{document}

\title{Hypothesis Testing in Feedforward Networks \\ with Broadcast Failures}

\author{Zhenliang~Zhang,~\IEEEmembership{Student~Member,~IEEE,}
        Edwin~K.~P.~Chong,~\IEEEmembership{Fellow,~IEEE,}\\
        Ali~Pezeshki,~\IEEEmembership{Member,~IEEE,}
        and
        William~Moran,~\IEEEmembership{Member,~IEEE}
\thanks{This work was supported in part by AFOSR under contract FA9550-09-1-0518, and by NSF under grants CCF-0916314 and CCF-1018472.}
\thanks{Z.~Zhang, E.~K.~P.~Chong, and A.~Pezeshki are with the Department of Electrical and Computer Engineering, Colorado State University, Fort Collins, CO 80523-1373, USA (e-mail: zhenliang.zhang@colostate.edu; edwin.chong@colostate.edu; ali.pezeshki@colostate.edu).
}
\thanks{W.~Moran is with the Department of Electrical and Electronic Engineering, The University of Melbourne, Melbourne, VIC 3010, Australia (e-mail: wmoran@unimelb.edu.au).
}
}

\maketitle

\begin{abstract}
Consider a large number of nodes, which sequentially make decisions between two given hypotheses. Each node takes a measurement of the underlying truth, observes the decisions from some immediate predecessors, and makes a decision between the given hypotheses. We consider two classes of broadcast failures: 1) each node broadcasts a decision to the other nodes, subject to random erasure in the form of a binary erasure channel; 2) each node broadcasts a randomly flipped decision to the other nodes in the form of a binary symmetric channel. We are interested in conditions under which there does (or does not) exist a decision strategy consisting of a sequence of likelihood ratio tests such that the node decisions converge in probability to the underlying
truth, as the number of nodes goes to infinity. In both cases, we
show that if each node only learns from a bounded number of
immediate predecessors, then there does not exist a decision
strategy such that the decisions converge in probability to the
underlying truth. However, in case 1, we show that if each node learns from an
unboundedly growing number of predecessors, then there exists a
decision strategy such that the decisions converge in probability to
the underlying truth, even when the erasure probabilities converge to
$1$. We show that a locally optimal strategy, consisting of a sequence
of Bayesian likelihood ratio tests, is such a strategy, and we derive the
convergence rate of the error probability for this strategy. In case
2, we show that if each node learns from all of its previous
predecessors, then there exists a decision strategy such that the decisions converge in probability to the
underlying truth when the flipping probabilities of the binary
symmetric channels are bounded away from $1/2$. Again, we show that a
locally optimal strategy achieves this, and we derive the convergence
rate of the error probability for it. In the case where the flipping
probabilities converge to $1/2$, we derive a necessary condition on the
convergence rate of the flipping probabilities such that the decisions
based on the locally optimal strategy still converge to the underlying
truth. We also explicitly characterize the relationship between the
convergence rate of the error probability and the convergence rate of
the flipping probabilities.
\end{abstract}


\section{Introduction}
We consider a large number of nodes, which sequentially make decisions between two hypotheses $H_0$ and $H_1$. At stage $k$, node $a_k$ takes a measurement $X_k$ (called its \emph{private signal}), receives the decisions of $m_k <k$ immediate predecessors, and makes a binary decision $d_k=0 \text{ or }1$ about the prevailing hypothesis $H_0$ or $H_1$, respectively. 
It then broadcasts a decision to its successors. Note that $m_k$ is often referred to as the \emph{memory size}. A typical question is this: Can these nodes asymptotically learn the underlying true hypothesis? In other words, does the decision $d_k$ converge (in probability) to the true hypothesis as $k\to \infty$? If so, what is the convergence rate of the error probability?

 One application of the sequential hypothesis testing problem\footnote{Our model for sequential hypothesis testing is different from
the model that goes by a similar name, due to Wald~\cite{wald}. In
Wald's sequential hypothesis testing problem, there is a single decision maker, who tests the given hypotheses by sequentially collecting samples. The sample size is not fixed in advance. Instead, according to the pre-defined stoping rule, the decision maker stops sampling and then declares a hypothesis.} is decentralized detection in sensor networks, in which case the set of nodes represents a set of spatially distributed sensors attempting to jointly solve the hypothesis testing problem, for example, the presence or absence of a target.  Decentralized detection problems have been intensively studied in recent years; see \cite{Var} for a comprehensive introduction to this problem. Usually, a sensor network consists of a large number of low-cost sensors with limited resources for processing and transmitting data. Therefore, each sensor has to aggregate its measurement and the observed messages from the previous sensors into a much smaller message (e.g., a 1-bit decision) and then sends it to other sensors for further aggregation. These sensors are subject to random failures, (e.g., dead battery), in which case the failed sensor cannot transmit its message. Moreover, the communication channels between sensors are noisy and the 1-bit messages are subject to random erasures or random flippings. A central question is whether or not there exists a sequence of decision rules for aggregating the spatially distributed information such that the decisions converge to the underlying truth as the number of sensors increases.

Another application is social learning in multi-agent networks, in which case the set of nodes represents a set of agents trying to learn the underlying truth (also known as the state of the world). Each agent makes a decision based on its own measurement and what it learns from the actions/decisions of the previous agents. In this case, we usually assume that each agent uses a myopic decision rule to minimize a local objective function; for example, the probability of error is locally minimized using the Bayesian likelihood ratio test with a threshold given by the ratio of the prior probabilities. The question in this setting is whether the agents in the social network can asymptotically learn the state of the world. 

To illustrate the feedforward nature of the model we study, consider
a customer having to decide whether or not to dine in a particular restaurant.
Typically, this decision is made based on her own taste and also
on the stated opinions of previous patrons. In this example, the
customer is a node in the feedforward network. The private signal at
this node represents the customer's own taste, while the received
decisions from predecessor nodes represent the perceived opinions of
previous patrons. Some previous patrons might not reveal their
opinions or might expose erroneous versions of their opinions. The
former is what we might call ``erasure'' of decisions, while the latter
represents ``flipping'' of decisions. We will formalize these
notions of erasure and flipping later. Similar examples along these
lines include customers deciding whether or not to watch a
particular movie and investors deciding whether or not to buy a
certain asset. A comprehensive exposition of social learning can be found in~\cite{Chris}.

\subsection{Related Work}
The literature on hypothesis testing in decentralized networks is vast, spanning various disciplines including signal processing, game theory, information theory, economics, biology, physics, computer science, and statistics. Here we only review the relevant asymptotic learning results in the network structure relevant to this paper.

The research on our problem begins with a seminal paper by Cover~\cite{Cover}, which considers the case where each node only observes the decision from its immediate previous node, i.e., $m_k=1$ for all $k$. This structure is also known as a serial network or tandem network and has been studied extensively in~\cite{Cover}--\nocite{Cover1,Swaszek,Tang,Tum,Kop,athans,Venu,Tsi,tandem,acemoglu1,tsi2,Var}\cite{Djuric}. We use $\mathbb P_j$ and $\pi_j$ to denote the probability measure and the prior probability associated with $H_j$, $j=0,1$, respectively. Cover~\cite{Cover} shows that if the (log)-likelihood ratio for each private signal $X_k$ is bounded almost surely, then using a sequence of likelihood ratio tests the (Bayesian) error probability \[\mathbb P_e^k=\pi_0 \mathbb P_0(d_k=1)+\pi_1 \mathbb P_1(d_k=0)\] does not converge in probability to 0 as $k\to\infty$. Conversely, if the likelihood ratio is unbounded, then the error probability converges to 0. In the case of unbounded likelihood ratios for the private signals, Veeravalli~\cite{Venu} shows that the error probability converges sub-exponentially with respect to the number $k$ of nodes in the case where the private signals are independent and follow identical Gaussian distribution. Tay \emph{et al.}~\cite{tandem} show that the convergence of error probability is in general sub-exponential and derive a lower bound for the convergence rate of the error probability in the tandem network.
Lobel \emph{et al.} \cite{acemoglu1} derive a lower bound for the convergence rate in the case where each node learns randomly from one previous node (not necessarily its immediate predecessor).
In the case of bounded likelihood ratios, Drakopoulos \emph{et al.}~\cite{tsi2} provide a non-Bayesian decision strategy, which leads to the convergence of the error probability.

Another extreme scenario is that each node can observe \emph{all} the previous decisions; i.e., $m_k=k-1$ for all $k$. This scenario was first studied in the context of social learning \cite{Banerjee}, \cite{Bik}, where each node uses the Bayesian likelihood ratio test to make its decision. In the case of bounded likelihood ratios for the private signals, the authors of \cite{Banerjee} and \cite{Bik} show that the error probability does not converge to 0, which results in arriving at the wrong decision with positive probability. 
In \cite{Zhang}, we show that in balanced binary trees, the decisions converge to the right decision even if the likelihood ratios of signals converge to 1 as the number of nodes increases. We further studied in  \cite{Zhang1} the convergence rate of the error probability in more general tree structures. In the case of unbounded likelihood ratios for the private signals,
Smith and Sorensen \cite{Smith} study this problem using martingales and show that the error probability converges to 0.
Krishnamurthy \cite{vicram0}, \cite{vicram} studies this problem from the perspective of quickest time change detection. Acemoglu \emph{et al.} \cite{acemoglu} show that the nodes can asymptotically learn the underlying truth in more general network structures.

Most previous work including those reviewed above assume that the nodes and links are perfect. We study the sequential hypothesis testing problem when broadcasts are subject to random erasure or random flipping.

\subsection{Contributions}
In this paper, we assume that each node uses a likelihood ratio test to generate its binary decision. We call the sequence of likelihood ratio tests a \emph{decision strategy}. We want to know whether or not there exists a decision strategy such that the node decisions converge in probability to the underlying true hypothesis. We consider two classes of broadcast failures:
\begin{itemize}
\item[1)] \emph{Random erasure}: Each broadcasted decision is erased with a certain erasure probability, modeled by a binary erasure channel. If the decision broadcasted by a node is erased, then none of its successors will observe that decision. 
\item[2)] \emph{Random flipping}: Each broadcasted decision is flipped with a certain flipping probability, modeled by a binary symmetric channel. If the broadcasted decision of a node is flipped, then all the successors of that node observe that flipped decision.

\end{itemize}

For case 1, we show that if each node can only learn from a bounded number of immediate predecessors, i.e., there exists a constant $C$ such that $m_k\leq C$ for all $k$, then for any decision strategy, the error probability cannot converge to 0.
We also show that if $m_k \to \infty$ as $k\to \infty$, then there exists a decision strategy such that the error probability converges to 0, even if the erasure probability converges to 1 (given that the convergence of the erasure probability is slower than a certain rate).
In the case where an agent learns from all its predecessors, the convergence rate of the error probability is $\Theta(1/\sqrt k)$. More specifically, we show that if the memory size $m_k=\Theta(k^{\sigma})$, $\sigma\leq 1$, then the error probability decreases as $\Theta(1/k^{\min{(\sigma,1/2)}})$.

For case 2, we show that if each node can only learn from a bounded number of immediate predecessors, then for any decision strategy, the error probability cannot converge to 0. We also show that if each node can learn from \emph{all} the previous nodes, i.e., $m_k=k-1$, then the error probability converges to 0 using the myopic decision strategy when the flipping probabilities are bounded away from $1/2$. In this case, we show that the error probability converges to 0 as $\Omega(1/k^{2})$. In the case where the flipping probability converges to $1/2$, we derive a necessary condition on the convergence rate of the flipping probability (i.e., how fast it must converge) such that the error probability converges to 0. More specifically, we show that if there exists $p>1$ such that the flipping probability converges to $1/2$ as $O(1/k(\log k)^p)$, then it is impossible that the error probability converges to 0. Therefore, only if  the flipping probability converges as $\Omega(1/k(\log k)^p)$ for some $p\leq 1$ can we hope for $\mathbb P_e^k\to 0$. Under this condition, we characterize explicitly the relationship between the convergence rate of the flipping probability and the convergence rate of the error probability.

\section{Preliminaries}
We use $\mathbb P$ to denote the underlying probability measure.
We use $\pi_j$ to denote the prior probability (assumed nonzero), $\mathbb P_j$ to denote the probability measure, and $\mathbb E_j$ to denote the conditional expectation associated with $H_j$, $j=0,1$. At stage $k$, node $a_k$ takes a measurement $X_k$ of the scene and makes a decision $d_k=0$ or $d_k=1$ about the prevailing hypothesis $H_0$ or $H_1$. It then broadcasts a potentially corrupted form $\hat d_k$ of that decision to its successors. Note that in case 1, if the decision is erased, it is equivalent
to saying that the corrupted decision $\hat d_k$ is $e$, which is a
message that carries no information and is not useful for
decision-making. Inserting $e$ in place of erased messages allows us
to unify the notation for cases 1 and 2. The decision $d_k$ of node $a_k$ is made based on the
private signal $X_k$ and the sequence of corrupted decisions $\hat D_{m_k}=\{\hat d_1, \hat d_2,\ldots,\hat d_{m_k}\}$ received from the $m_k$ immediate predecessor
nodes using a likelihood ratio test.

Our aim is to find a sequence of likelihood ratio tests such that  the probability of
making a wrong decision about the state of the world tends to $0$ as
$k\to\infty$; i.e., \[
\lim_{k\to\infty}\mathbb P_e^k=\lim_{k\to\infty}(\pi_0 \mathbb P_0(d_k=1)+\pi_1 \mathbb P_1(d_k=0) )= 0.\]
Before proceeding, we introduce the following definitions and assumptions:

\begin{enumerate}

\item The private signal $X_k$ takes values in a set $S$, endowed with a $\sigma$-algebra $\mathcal S$. We assume that $X_k$ is independent of the broadcast history
$\hat D_{m_k}$. 
Moreover, the $X_k$s are mutually independent and identically distributed
with distribution $\mathbb P_j^X$, under $H_j$, $j=0,1$. (Note that $\mathbb P_j^X$ is a probability measure on the $\sigma$-algebra $\mathcal S$.) We assume that the underlying
hypothesis, $H_0$ or $H_1$, does not change with $k$.

\item The two probability measures $\mathbb P_0^X$ and $\mathbb P_1^X$ are equivalent; i.e., they are absolutely continuous with respect to each other. In other words, if $A\in \mathcal S$, then $\mathbb P_0^X(A)=0$ if and only if $\mathbb P_1^X(A)=0$.

\item Let the likelihood ratio of a private signal $s\in S$ be \[
L_X(s)=\frac{d\mathbb P_1^X}{d\mathbb P_0^X}(s),\]
where $d\mathbb P_1^X/d\mathbb P_0^X$ denotes the Radon--Nikodym derivative (which is guaranteed to exist because of the assumption that the two measures are equivalent).
We assume that the likelihood ratios for the private signals are unbounded; i.e., for any set $S' \subset S$ with probability 1 under the measure $(\mathbb P_0^X+\mathbb P_1^X)/2$, we have
\[\inf_{s\in S'} \frac{d \mathbb P_1^X}{d\mathbb P_0^X}(s)=0\] and
\[\sup_{s\in S'} \frac{d \mathbb P_1^X}{d\mathbb P_0^X}(s)=\infty.\]

\item
Suppose that $\theta$ is the underlying truth.
Let $\bar b_k=\mathbb P(\theta=H_1| X_k),$ which we call the \emph{private belief} of $a_k$.
By Bayes' rule, we have
\begin{align}
\bar b_k
=\left(1+ \frac{\pi_0}{\pi_1} \frac{1}{L_X(X_k)}  \right)^{-1}.
\label{eq:mu}
\end{align}

\item Recall that node $a_k$ observes $m_k$ decisions $\hat D_{m_k}$ from its immediate predecessors. Let $p^k_j$ be the conditional probability mass function of $\hat D_{m_k}$ under $H_j$, $j=0,1$.  The likelihood ratio of a realization $\mathcal D_{m_k}$ is
\[
L_D^k(\mathcal D_{m_k})=\frac{p_1^k(\mathcal D_{m_k})}{p_0^k(\mathcal D_{m_k})}=\frac{\mathbb P_1(\hat D_{m_k}=\mathcal D_{m_k})}{\mathbb P_0(\hat D_{m_k}=\mathcal D_{m_k})}.
\]

\item Let $b_k =\mathbb P(\theta=H_1| \hat D_{m_k}),$ which we call the \emph{public belief} of $a_k$. We have
\begin{align}
 b_k=\left(1+ \frac{\pi_0}{\pi_1} \frac{1}{L_D^k(\hat D_{m_k})}  \right)^{-1}.
\label{eq:pb}
\end{align}

\item Each node $a_k$ makes its decision using its own measurement and the observed decisions based on a likelihood ratio test with a threshold $t_k>0$:
\begin{equation*}
d_k=
\begin{cases} 1 & \text{ if } L_X(X_k)L_D^k(\hat D_{m_k})> t_k,
\\
0 & \text{ if } L_X(X_k)L_D^k(\hat D_{m_k})\leq t_k.
\end{cases}
\end{equation*}

If $t_k=\pi_0/\pi_1$, then this test becomes the maximum a-posteriori probability (MAP) test, in which case the probability of error is locally minimized for node $a_k$. If $t_k=1$, then the test becomes the maximum-likelihood (ML) test. If the prior probabilities are equal, then these two tests are identical.
A decision strategy $\mathbb T$ is a sequence of likelihood ratio tests with thresholds $\{t_k\}_{k=1}^{\infty}$. Given a decision strategy, the decision sequence $\{d_k\}_{k=1}^{\infty}$ is a well-defined stochastic process.

\item We say that the system \emph{asymptotically learns} the underlying true hypothesis with decision strategy $\mathbb T$ if 
\[
\lim_{k\to\infty} \mathbb P(d_k=\theta) =1.
\]
In other words, the probability of making a wrong decision goes to 0, i.e., $\lim_{k\to\infty}\mathbb P_e^k =0$. The question we are interested in is this: In each of the two classes of failures, is there a decision strategy such that the system asymptotically learns the underlying true hypothesis? 
\end{enumerate}

\section{Random Erasure}
In this section, we consider the sequential hypothesis testing problem in the presence of random erasures, modeled by binary erasure channels. Recall that the binary message $d_k$ is the input to a binary erasure channel and $\hat d_k$ is the output, which is either equal to $d_k$ (no erasure) or is equal to a symbol $e$ that represents the occurrence of an erasure. The erasure channel matrix at stage $k$ is given by  $\mathbb P(\hat d_k=i| d_k=j),$ $j=0,1 \text{ and } i=j,e$. Recall that each node $a_k$ observes $m_k$ immediate previous broadcasted decisions.
We divide our analysis into two scenarios: \emph A) $\{m_k\}$ is bounded above by a positive constant; \emph B) $m_k$ goes to infinity as $k\to \infty$.

\subsection{Bounded Memory}

\begin{Theorem} Suppose that there exists $C$ and $\epsilon>0$ such that for all $k$, $m_k\leq C$ and $\mathbb P(\hat d_k=e| d_k=j) \in [\epsilon,1-\epsilon]$ for $j=0,1$. Then, there does not exist a decision strategy such that the error probability converges to 0.
\end{Theorem}

\begin{IEEEproof} We first prove this claim for the special case of the {tandem network}, where $m_k=1$ for all $k$. For each node $a_k$, with a nonzero probability $\mathbb P(\hat d_k=e| d_k=j)$, the decision $d_{k-1}=j$ of the immediate predecessor is erased and $a_k$ makes a decision based only on its own private signal $X_k$. We use $\mathcal E_k$ to denote this event. Conditioned on $\mathcal E_k$, we claim that the error probability as a sequence of~$k$,
\begin{align*}
&\mathbb P(d_k\neq \theta|\mathcal E_k) \\
&=\pi_0 \mathbb P_0(d_k=1|\mathcal E_k)+\pi_1 \mathbb P_1(d_k=0|\mathcal E_k)\\
&= \pi_0 \mathbb P_0(L_X(X_k)> t_k)+\pi_1 \mathbb P_1(L_X(X_k)\leq t_k),
\end{align*}
 is bounded away from 0.
We prove the above claim by contradiction.
Suppose that there exists a decision strategy with threshold sequence $\{t_k\}$ such that $\mathbb P(d_k\neq \theta|\mathcal E_k)\to 0$ as $k\to\infty$. Then, we must have $\mathbb P_1(L_X(X_k)\leq t_k)\to 0$ because $\pi_1$ is positive. Because $\mathbb P_0^X$ and $\mathbb P_1^X$ are equivalent measures, we have $\mathbb P_0(L_X(X_k)\leq t_k) \to 0$. Hence we have $\mathbb P_0(L_X(X_k)> t_k) \to 1$. Therefore, $\mathbb P(d_k\neq \theta|\mathcal E_k)$ does not converge to 0. 

We use $\mathcal E_k^{\mathcal C}$ to denote the complement event of $\mathcal E_k$. By the Law of Total Probability, we have \begin{align*}
\mathbb P_e^k & =\mathbb P(\mathcal E_k) \mathbb P(d_k\neq \theta|\mathcal E_k)+\mathbb P(\mathcal E_k^{\mathcal C}) \mathbb P(d_k\neq \theta|\mathcal E_k^{\mathcal C}) \\
&\geq \mathbb P(\mathcal E_k) \mathbb P(d_k\neq \theta|\mathcal E_k).\end{align*} Because $\mathbb P(\mathcal E_k)\geq \epsilon$, we conclude that the error probability does not converge to 0.

We can now generalize this proof to the case of a general bounded $m_k$ sequence. Let $\mathcal E_k$ be the event that $a_k$ receives $m_k$ erased symbols $e$. Then, the probability $\mathbb P(\mathcal E_k)$ is bounded below according to \[
\mathbb P(\mathcal E_k) \geq \left(\min_{\substack{j=0,1\\
		m=k-1,\ldots,k-m_k
                   }}\mathbb P(\hat d_m=e|d_m=j)\right)^{m_k}\geq \epsilon^{m_k}.\]
We have already shown that given this event the error probability does not converge to 0. Using the Law of Total Probability, It is easy to see that the error probability does not converge to 0.  \end{IEEEproof}

\begin{rem} 
We use $\mathbb P(\hat d_k=e| d_k=j) \in [\epsilon,1-\epsilon]$ for $j=0,1$ to mean that the erasure probability $\mathbb P(\hat d_k=e| d_k=j)$ is bounded away from 0 and 1.
\end{rem}

This result is straightforward to understand. If the memory sizes are bounded for all nodes, then for each node, there exists a positive probability such that all the decisions received from its immediate predecessors are erased, in which case the node has to make a decision based on its own measurement. The error probability cannot converge to 0 because of the  equivalent-measure assumption.

\subsection{Unbounded Memory}
Suppose that each node $a_k$ observes $m_k$ immediate previous decisions. In this section, we deal with the case where $m_k$ is unbounded.\footnote{The assumption that $m_k$ is unbounded is not sufficiently strong to guarantee the convergence of error probability to 0. An example is that the memory size $m_k$ equals $\sqrt k$ if $\sqrt k$ is an integer and it equals $1$ otherwise. In this case, we can use a similar argument as that in the proof of Theorem~1 to show that the error probability does not converge to 0. } More specifically, we consider the case where $m_k$ goes to infinity. We first consider the case where the erasure probabilities are bounded away from 1. We have the following result.

\begin{Theorem} Suppose that $m_k$ goes to infinity as $k\to\infty$ and there exists $\epsilon>0$ such that for all $j=0,1$ and for all $k$, $\mathbb P(\hat d_k=e|d_k=j)\leq 1-\epsilon$.
Then, there exists a decision strategy such that the error probability converges to 0.

\end{Theorem}
\begin{IEEEproof}
We prove this result by constructing a certain tandem network within the original network using a \emph{backward-searching scheme}. The scheme is the following: Consider node $a_k$ in the original network. Let $n_k$ be the largest integer such that each node in the sequence $\{a_{k-n_k^2}, a_{k-n_k^2-1}, \ldots, a_{k}\}$ of $n_k^2+1$ nodes has a memory size that is greater than or equal to $n_k$. Note that an $n_k$ satisfying this condition is guaranteed to exist. Moreover, because $m_k$ goes to infinity as $k\to\infty$, we have $n_k \to \infty$ as $k\to\infty$. Consider the event that $a_{k}$ receives at least one decision $j$, which is not erased, from $\{a_{k-n_k},\ldots, a_{k-1}\}$, its $n_k$ immediate predecessors. The probability of this event is at least \[1-\max_{\substack{j=0,1\\
		m=k-n_k,\ldots, k-1
                   }}\mathbb P(\hat d_m=e|d_m=j)^{n_k},\] 
which is bounded below by $1-(1-\epsilon)^{n_k}$ by the assumption on the erasure probabilities. We denote the node that sends the unerased decision by $a_{k_1}$. Similarly, with a certain probability, $a_{k_{1}}$ receives at least one decision, which is not erased, from its $n_k$ immediate predecessors. Recursively, with a certain probability, we can construct a tandem network with length $n_k$ using nodes from among the $n_k^2+1$ nodes above within the original network. Let $\mathcal E_{k}$ be the event that such a tandem network exists. The probability $\mathbb P(\mathcal E_{k})$ is at least $(1-(1-\epsilon)^{n_k})^{n_k}$. Recall that $\lim_{k\to\infty} n_k =\infty$, which implies that \[\lim_{k\to\infty} (1-(1-\epsilon)^{n_k})^{n_k}=1.\] Hence we have \[\lim_{k\to\infty} \mathbb P(\mathcal E_{k}) =1.\] Conditioned on $\mathcal E_k$, by using the strategy $\mathbb T$ consisting of a sequence of likelihood ratio tests with monotone thresholds described in \cite{Cover}, we can get the conditional convergence of the error probability, given $\mathcal E_k$, to 0. We can also use the equilibrium strategy described in \cite{acemoglu1}. Therefore, by the Law of Total Probability, we have
\begin{align}
\nonumber
&\lim_{k\to \infty}\mathbb P(d_k\neq\theta)\\
\nonumber
&=\lim_{k\to\infty}\left(\mathbb P(d_k\neq\theta|\mathcal E_{k}) \mathbb P(\mathcal E_{k})+\mathbb P(d_k\neq \theta|\mathcal E_{k}^{\mathcal C}) (1-\mathbb P(\mathcal E_{k})\right) \\
& \leq \lim_{k\to\infty} \left(\mathbb P(d_k\neq\theta|\mathcal E_{k}) +(1-\mathbb P(\mathcal E_{k})\right) =0.
\label{eq:rate}
\end{align}
\end{IEEEproof}

Note that given a strategy, the convergence rate for the error probability in this case depends on how fast $\mathbb P(\mathcal E_{k})$ converges to 1 and how fast $\mathbb P(d_k\neq\theta|\mathcal E_{k})$ converges to 0.

First let us consider the convergence rate of $\mathbb P(\mathcal E_{k})$. Obviously this convergence rate depends on the convergence rate of $n_k$. Moreover, the convergence rate of $n_k$ depends on the convergence rate of $m_k$. 
For example, if $m_k$ goes to infinity extremely slowly, then $n_k$ grows extremely slowly with respect to $k$, which means that $\mathbb P(\mathcal E_{k})$ converges to 1 extremely slowly with respect to $k$.
Next we assume that $m_k$ increases as $\Theta(k^{\sigma})$, where $\sigma \leq 1$.
We first establish a relationship between the convergence rate of $m_k$ and the convergence rate of $n_k$ when using the backward-searching scheme.

\begin{Prop} Suppose that $m_k=\Theta(k^{\sigma})$ where $\sigma\leq 1$. Then, we have
\begin{equation*}
n_k=
\begin{cases} \Theta(\sqrt k) & \text{ if } \sigma \geq 1/2,
\\
\Theta(k^{\sigma}) & \text{ if } \sigma<1/2.
\end{cases}
\end{equation*}

\end{Prop}
\begin{IEEEproof} Suppose that we can form a tandem network with length $n_k$ within the original network. Recall that $n_k$ is the largest integer such that each node in the sequence $\{a_{k-n_k^2}, a_{k-n_k^2-1}, \ldots, a_{k}\}$ of $n_k^2+1$ nodes has a memory size that is greater than or equal to $n_k$. Therefore, the memory size $m_{k-n_k^2}$ of $a_{k-n_k^2}$ must be larger than or equal to $n_k$ by assumption. Hence we have
\begin{align*}
m_{k-n_k^2}=(k-n_k^2)^{\sigma} \geq n_k. 
\end{align*}
 Moreover, the memory size $m_{k-(n_k+1)^2}$ of $a_{k-(n_k+1)^2}$ must be strictly smaller than $n_k+1$ (otherwise we can construct a tandem network with length $n_k+1$). Hence we have
\begin{align*}
m_{k-(n_k+1)^2}=(k-(n_k+1)^2)^{\sigma} < n_k+1.
\end{align*}
From the above two inequalities, we easily obtain the desired asymptotic rates for $n_k$.
\end{IEEEproof}

\begin{rem} Note that if $\sigma<1/2$, then the scaling law of $n_k$ is identical to that of $m_k$: The faster the scaling of $m_k$, the faster the scaling of $n_k$ also. However, for $\sigma\geq 1/2$, the scaling law of $n_k$ ``saturates'' at $\sqrt{k}$, no matter how fast $m_k$ scales.
\end{rem}
We have derived the convergence rate for $n_k$.
Recall that $\mathbb P(\mathcal E_{k})$ converges to 1 at least in the rate of $\Theta(n_k (1-\epsilon)^{n_k})$. From this fact and Proposition 1, we derive the convergence rate for $\mathbb P(\mathcal E_{k})$.
\begin{Cor} Suppose that $m_k=\Theta(k^{\sigma})$ where $\sigma\leq 1$. Then, we have
\begin{equation*}
1-\mathbb P(\mathcal E_{k})=
\begin{cases} O(\sqrt k (1-\epsilon)^{\sqrt k}) & \text{ if } \sigma \geq 1/2,
\\
O( k^{\sigma} (1-\epsilon)^{ k^{\sigma}}) & \text{ if } \sigma<1/2.
\end{cases}
\end{equation*}

\end{Cor}
\bigskip

Second, let us consider the convergence rate of $\mathbb P(d_k\neq\theta|\mathcal E_{k})$. Recall that $\mathcal E_{k}$ denotes the event that a tandem network with length $n_k$ exists. Conditioned on $\mathcal E_{k}$, if we use the the equilibrium strategy\footnote{Note that this equilibrium strategy is \emph{not} the only strategy such that the error probability converges to 0 in a tandem network.} described in \cite{acemoglu1}, then it has been shown that the error probability converges to 0 as $\Theta(1/n_k)$, with appropriate assumptions on the distributions of the private signal. From this fact and Proposition 1, we derive the convergence rate for $\mathbb P(d_k\neq\theta|\mathcal E_{k})$.
\begin{Cor} Suppose that $m_k=\Theta(k^{\sigma})$ where $\sigma\leq 1$. Then, we have
\begin{equation*}
\mathbb P(d_k\neq\theta|\mathcal E_{k})=
\begin{cases} \Theta(1/\sqrt k) & \text{ if } \sigma \geq 1/2,
\\
\Theta(1/k^{\sigma}) & \text{ if } \sigma<1/2.
\end{cases}
\end{equation*}

\end{Cor}
\bigskip

Notice that the convergence rate of $\mathbb P(d_k\neq\theta|\mathcal E_{k})$ is much smaller than that of $\mathbb P(\mathcal E_{k})$. Moreover by~\eqref{eq:rate}, the convergence rate of $\mathbb P(d_k\neq\theta)$ depends on the smaller of the convergence rates of $\mathbb P(d_k\neq\theta|\mathcal E_{k})$ and $\mathbb P(\mathcal E_{k})$. We derive the convergence rate for the error probability as follows.
\begin{Cor} Suppose that $m_k=\Theta(k^\sigma)$ where $\sigma\leq 1$. Then, we have
\begin{equation*}
\mathbb P(d_k\neq\theta)=
\begin{cases} \Theta(1/\sqrt k) & \text{ if } \sigma \geq 1/2,
\\
\Theta(1/k^{\sigma}) & \text{ if } \sigma<1/2.
\end{cases}
\end{equation*}

\end{Cor}

\bigskip

We have considered the situation where the erasure probabilities are bounded away from 1. Now consider the case where the erasure probability $\mathbb P(\hat d_k=e|d_k=j)$ converges to~1.
\begin{Theorem}
Suppose that $\mathbb P(\hat d_k=e|d_k=j)\to 1$ and there exists $\epsilon >1$ and $c>0$ such that $\mathbb P(\hat d_k=e|d_k=j)  \leq (c n_k )^{-{\epsilon}/{n_k}}$. Then, there exists a decision strategy such that the error probability converges to 0.

\end{Theorem}
\begin{IEEEproof}
We use the scheme described in the proof of Theorem 2. The probability that a tandem network with length $n_k$ exists is at least $(1-((cn_k)^{-{\epsilon}/{n_k}})^{n_k})^{n_k}=(1-(cn_k)^{-\epsilon})^{n_k}$, which converges to 1 as $k\to\infty$. Using the same arguments as those in the proof of Theorem~2, we can show that the error probability converges to 0.
\end{IEEEproof}

As an example, we consider the situation where each node observes \emph{all} the previous decisions; i.e, $m_k=k-1$ for all $k$. In this case, it is easy to show that using the backward-searching scheme, with a certain probability, we can form a tandem network with length $n_k=\lfloor \sqrt {k-1} \rfloor$. 
Suppose that the erasure probabilities are bounded away from 1. Then, the error probability converges to 0 as $\Theta(1/\sqrt k)$. Moreover, the error probability converges to 0 even if the erasure probability converges to 1, provided that $\mathbb P(\hat d_k=e|d_k=j)  \leq (c n_k )^{-{\epsilon}/{n_k}}$.

\section{Random Flipping}

We study in this section the sequential hypothesis testing problem with random flipping, modeled by a binary symmetric channel. 
Recall that $d_k$ is the input to a binary symmetric channel and $\hat d_k$ is the output, which is either equal to $d_k$ (no flipping) or is equal to its complement $1-d_k$ (flipping). The channel matrix is given by
$\mathbb P(\hat d_k=i| d_k=j),$ $i,j=0,1$. We assume that $\mathbb P(\hat d_k=1| d_k=0)=\mathbb P(\hat d_k=0| d_k=1)=q_k$, where $q_k$ denotes the probability of a flip. The assumption of symmetry is for simplicity only, and all results obtained in this section can be generalized easily to a general binary communication channel with unequal flipping probabilities, i.e., $\mathbb P(\hat d_k=1| d_k=0)\neq \mathbb P(\hat d_k=0| d_k=1)$.
We assume that each node $a_k$ knows the probabilities of flipping associated with the corrupted decisions $\hat D_{m_k}$ received from its predecessors.

\subsection{Bounded Memory}

\begin{Theorem} Suppose that there exists $C$ and $\epsilon>0$ such that for all $k$, $m_k \leq C$ and $q_k\in [\epsilon,1-\epsilon]$. Then, there does not exist a decision strategy such that the error probability converges to 0.

\end{Theorem}

\begin{IEEEproof} We first prove this theorem in the case where each node observes the immediate previous node; i.e., $m_k=1$ for all $k$. Node $a_k$ makes a decision $d_k$ based on its private signal $X_k$ and the decision $\hat d_{k-1}$ from its immediate predecessor. Recall that $q_k=\mathbb P(\hat d_k=1| d_k=0)=\mathbb P(\hat d_k=0| d_k=1)$. The likelihood ratio test at stage $k$ (with a threshold $t_k> 0$) is
\begin{equation*}
d_k=
\begin{cases} 1 & \text{ if } L_X(X_k)L_D^k(\hat d_{k-1})> t_k,
\\
0 & \text{ if } L_X(X_k)L_D^k(\hat d_{k-1})\leq t_k,
\end{cases}
\end{equation*}
where for each $j_{k-1}=0,1$
\[
L_D^k(j_{k-1})=\frac{p_1^k(j_{k-1})}{p_0^k(j_{k-1})}=\frac{\mathbb P_1(\hat d_{k-1}=j_{k-1})}{\mathbb P_0(\hat d_{k-1}=j_{k-1})},
\]
and $\mathbb P_j(\hat d_{k-1}=j_{k-1})$, $j=0,1$ is given by
\begin{align}
\label{eq:pp}\nonumber
\mathbb P_{j}(\hat d_{k-1}=j_{k-1})&=q_k(1-\mathbb P_j(d_{k-1}=j_{k-1}))\\
\nonumber
&\quad+(1-q_k)\mathbb P_j(d_{k-1}=j_{k-1})\\
&=q_k+(1-2q_k)\mathbb P_j(d_{k-1}=j_{k-1}).\end{align}

Let $t_k(\hat d_{k-1})=t_k/L_D^k(\hat d_{k-1})$ be the testing threshold for $L_X(X_k)$ when $\hat d_{k-1}$ is received. Then, the likelihood ratio test can be rewritten as
\begin{equation*}
d_k=
\begin{cases} 1 & \text{ if } L_X(X_k)> t(\hat d_{k-1}),
\\
0 & \text{ if } L_X(X_k)\leq t(\hat d_{k-1}).
\end{cases}
\end{equation*}
From \eqref{eq:pp}, we notice that $\mathbb P_j(\hat d_{k-1})$ depends linearly on $\mathbb P_j(d_{k-1})$.
Without loss of generality, henceforth we assume that $q_k \leq 1 /2$.\footnote{Note that the system is symmetric with respect to $q_k=1/2$. For example, if the probability of flipping is 1, i.e., $q_k=1$, then the receiver can revert the received decision back since it knows the predecessor always `lies.' }
It is obvious that $t_k(0)\geq t_k(1)$ because $L_D^k(j)=\mathbb P_1(\hat d_{k-1}=j)/\mathbb P_0(\hat d_{k-1}=j)$ is non-decreasing in $j$. Therefore, the likelihood ratio test becomes
\begin{equation*}
d_k=
\begin{cases} 1 & \text{ if } L_X(X_k) > t_k(0),
\\
0 & \text{ if } L_X(X_k)\leq t_k(1),
\\
\hat d_{k-1} & \text{otherwise,}
\end{cases}
\end{equation*}
and we can write the Type~I and Type~II error probabilities, denoted by $\mathbb P_0(d_k=1)$ and $\mathbb P_1(d_k=0)$, respectively, as follows:
\begin{align*}
\mathbb P_0(d_k=1)&=\mathbb P_0(L_X(X_k)> t_k(0))\mathbb P_0(\hat d_{k-1}=0)\\
&\quad+\mathbb P_0(L_X(X_k)> t_k(1))\mathbb P_0(\hat d_{k-1}=1)
\end{align*}
and
\begin{align*}
\mathbb P_1(d_k=0)&=\mathbb P_1(L_X(X_k)\leq t_k(0))\mathbb P_1(\hat d_{k-1}=0)\\
&\quad+\mathbb P_1(L_X(X_k)\leq t_k(1))\mathbb P_1(\hat d_{k-1}=1).
\end{align*}
The total error probability at stage $k$ is
\begin{align*}
\mathbb P_e^k &=  \pi_0 \mathbb P_0(d_k=1)+\pi_1 \mathbb P_1(d_k=0)\\
&= \pi_0(\mathbb P_0(L_X(X_k)> t_k(0))\\
&\quad+\mathbb P_0(t_k(1) < L_X(X_k)\leq t_k(0))\mathbb P_0(\hat d_{k-1}=1)) \\
&\quad+\pi_1(\mathbb P_1(t_k(1)< L_X(X_k)\leq t_k(0))\mathbb P_1(\hat d_{k-1}=0)\\
&\quad+\mathbb P_1(L_X(X_k)\leq t_k(1))).
\end{align*}

We prove the claim by contradiction.
Suppose that there exists a strategy such that $\mathbb P_e^k\to 0$ as $k\to \infty$. Then, we must have $\mathbb P_0(L_X(X_k)> t_k(0))\to 0$ and $\mathbb P_1(L_X(X_k)\leq t_k(1)) \to 0$. Recall that $\mathbb P_0^X$ and $\mathbb P_1^X$ are equivalent measures. Hence we have $\mathbb P_1(L_X(X_k)> t_k(0))\to 0$ and $\mathbb P_0(L_X(X_k)\leq t_k(1)) \to 0$. These imply that $\mathbb P_j(t_k(1) < L_X(X_k)\leq t_k(0))\to 1$ for $j=0,1$. But 
\begin{align*}\mathbb P_j(\hat d_{k-1}=1-j)&=q_{k}(1-\mathbb P_j(d_{k-1}=1-j))\\
&\quad+(1-q_k)\mathbb P_j(d_{k-1}=1-j)\\
&=q_k+(1-2q_k)\mathbb P_j(d_{k-1}=1-j),\end{align*}
which is bounded below by $q_k$. Hence $\mathbb P_e^k$ is also bounded below away from 0 in the asymptotic regime. This contradiction implies that $\mathbb P_e^k$ does not converge to 0. The proof for the general bounded memory case is similar and is given in Appendix A.
\end{IEEEproof}

\subsection{Unbounded Memory}
In this section, we consider the case where $a_k$ can observe all its predecessors; i.e., $m_k=k-1$. We will show that using the myopic decision strategy, the error probability converges to 0 in the presence of random flipping when the flipping probabilities are bounded away from $1/2$. In the case where the flipping probability converges to $1/2$, we derive a necessary condition on the convergence rate of the flipping probability such that the error probability converges to 0. Moreover, we precisely describe the relationship between the convergence rate of the flipping probability and the convergence rate of the error probability.

 If we state the conditions on the
private signal distributions in a symmetric way, then it suffices to consider the case
when the true hypothesis is $H_0$. In this case, our aim is to show that the Type~I error probability converges to 0, i.e., $\mathbb P_0(d_k=1)\to 0$. We consider the myopic decision strategy; i.e., the decision made by the $k$th node is on the basis of the MAP test.
Again, the corruption from $d_k$ to $\hat d_k$ is in the form of a binary symmetric channel with flipping probability denoted by $q_k$. Without loss of generality, we assume that $q_k\leq 1/2$ (because of symmetry). 
We define the \emph{public likelihood ratio} of $\mathcal D_k=(j_1,j_2,\ldots,j_k)$ to be \[
L_k(\mathcal D_{k})=\frac{p_1^k(\mathcal D_{k})}{p_0^k(\mathcal D_{k})}=\frac{\mathbb P_1(\hat D_{k}=\mathcal D_{k})}{\mathbb P_0(\hat D_{k}=\mathcal D_{k})}.
\]
We will consider two cases:
\begin{itemize}
\item[1)] The flipping probabilities are bounded away from $1/2$ for all $k$; i.e., there exists $c>0$ such that $q_k \leq 1/2- c$ for all $k$. This ensures that the corrupted decision still contains some useful information about the true hypothesis. We call this the case of \emph{uniformly informative nodes}.
\item[2)] The flipping probabilities $q_k$ converge to $1/2$; i.e., $q_k\to1/2$ as $k\to \infty$. This means that the broadcasted decisions become increasingly uninformative as we move towards the latter nodes. We call this the case of \emph{asymptotically uninformative nodes}.
 \end{itemize}

\subsubsection{Uniformly informative nodes}
We first show that the error probability converges to 0.
Recall that $\bar b_k=\mathbb P(\theta=H_1|X_k)$ denotes the private belief given by signal $X_k$.
Let $(\mathbb G_0, \mathbb G_1)$ be the conditional distributions of the private belief $\bar b_k$:
\[
\mathbb G_j(r)=\mathbb P_j(\bar b_k \leq r).
\]
Note that $\mathbb G_j$ does not depend on $k$ because the $X_k$s are identically distributed. 
These distributions exhibit two important properties:
\begin{enumerate}
\item[a)] \emph{Proportionality}: This property is easy to get from Bayes' rule: for all $r\in (0,1)$, we have
\[
\frac{d \mathbb G_1}{d \mathbb G_0} (r)=\frac{r}{1-r},
\]
where $d\mathbb G_1/d \mathbb G_0$ is the Radon-Nikodym derivative of their associated probability measures.
\item[b)] \emph{Dominance}: $\mathbb G_1(r)<\mathbb G_0(r)$ for all $r\in (0,1)$, and $\mathbb G_j(0) =0 \text{ and } \mathbb G_j(1)=1$ for $j=0,1$. Moreover, $\mathbb G_1(r)/\mathbb G_0(r)$ is monotone non-decreasing as a function of $r.$
\end{enumerate}

We define an increasing  sequence $\{\mathcal F_k\}$ of $\sigma$-algebras as
follows:
\begin{equation*}
  \mathcal F_k=\sigma\langle X_1,X_2,\ldots, X_k;  \hat d_1, \hat d_2,\ldots, \hat d_k \rangle.
\end{equation*}
Evidently $\hat d_k$ and $L_k(\hat D_k)$ are adapted to  this sequence of $\sigma$-algebras. Moreover, given $\hat D_{k-1} =\{\hat d_1, \hat d_2,\ldots, \hat d_{k-1}\}$ and $X_k$, the decision $d_k$ is completely determined. Therefore, $d_k$ is also adapted to this sequence of $\sigma$-algebras.

%
%

\begin{lemma}
  \label{thm:1}
 Under hypothesis  $H_0$, the public
  likelihood ratio sequence $\{L_k(\hat D_k)\}$ is a martingale with respect to
  $\{\mathcal F_k\}$ and $L_k(\hat D_k)$ converges to a finite limit almost surely.
\end{lemma}
\begin{IEEEproof}
 The expectation of $L_{k+1}(\hat D_{k+1})$
  conditioned on $H_0$ and $\mathcal F_k$  is
  \begin{align*}
&\mathbb E_0[L_{k+1}(\hat D_{k+1})|\mathcal F_k]=\sum_{\hat d_{k+1}=0,1} \mathbb P_0(\hat d_{k+1}|\mathcal F_k) L_{k+1}(\hat D_{k+1}) \\
&=\sum_{\hat d_{k+1}=0,1} \mathbb P_0(\hat d_{k+1}|\mathcal F_k) L_{k}(\hat D_k) \frac{\mathbb P_1(\hat d_{k+1}|\mathcal F_k)}{\mathbb P_0(\hat d_{k+1}|\mathcal F_k)} \\
&=L_k(\hat D_k) \sum_{\hat d_{k+1}=0,1} \mathbb P_0(\hat d_{k+1}|\mathcal F_k)  \frac{\mathbb P_1(\hat d_{k+1}|\mathcal F_k)}{\mathbb P_0(\hat d_{k+1}|\mathcal F_k)} \\
&=L_k(\hat D_k).
\end{align*}

Moreover, note that 
\[
\int |L_1(\hat D_1)| \text{ }d \mathbb P_0 =1 <\infty.
\]
Since $L_k(\hat D_k)$ a non-negative martingale, by Doob's martingale convergence theorem \cite{pro}, it converges almost surely  to a finite limit.
\end{IEEEproof}

Let $L_\infty$ be the almost sure limit of
$L_k(\hat D_k)$ conditioned on $H_0$, and note that
$L_\infty<\infty$ almost surely. This claim holds for both cases 1 and 2.
By \eqref{eq:pb}, we know that the public belief $b_k<1$ almost surely. The implication is that the public belief cannot go completely wrong.
Moreover, for case 1, we can show that the public likelihood ratio converges to 0 almost surely.

\begin{lemma}   \label{thm:2}
Suppose that the flipping probabilities are bounded away from $1/2$. Then under $H_0$, we have
  $L_{\infty}=0$ almost surely.
\end{lemma}
\begin{IEEEproof}
For the public likelihood ratio, we have the following recursion:
  \begin{align}
  \nonumber
    L_{k+1}(\hat D_{k+1})&=\frac{\mathbb P_1(\hat D_{k+1})}{\mathbb  P_0(\hat D_{k+1})}\\
    &=
      \frac{\mathbb P_1(\hat d_{k+1}|\hat D_{k})}{\mathbb P_0(\hat d_{k+1}|\hat
        D_{k})}L_{k}(\hat D_{k}).
            \label{eq:7}
  \end{align}
  Consider the event $A=\{L_{\infty}>0\}$. 
On $A$, we have
   \begin{equation}
    \label{eq:14}
    \frac{\mathbb P_1(\hat d_{k+1}|\hat D_{k})}{\mathbb P_0(\hat d_{k+1}|\hat
        D_{k})}\to 1,
  \end{equation}
  almost everywhere.
Now 
\begin{align}
\nonumber
  \frac{\mathbb P_1(\hat d_{k+1}|\hat D_{k})}{\mathbb P_0(\hat d_{k+1}|\hat
  D_{k})} & =  \frac{ \sum_{ d_{k+1}}\mathbb P_1( d_{k+1}|\hat
  D_{k})\mathbb P(\hat d_{k+1}|d_{k+1})}{ \sum_{ d_{k+1}}\mathbb P_0( d_{k+1}|\hat
  D_{k})\mathbb P(\hat d_{k+1}|d_{k+1})}\\
&=\frac{\mathbb P_1( d_{k+1}|\hat
  D_{k})(1-2q_k)+q_k}
{\mathbb P_0( d_{k+1}|\hat D_{k})(1-2q_k)+q_k}.\label{eq:15}
\end{align}
Equation~\eqref{eq:15} together with \eqref{eq:14}
implies
\begin{equation*}
\frac{\mathbb P_1(d_{k+1}|\hat D_{k})}{\mathbb P_0( d_{k+1}|\hat
  D_{k})}\to 1,
\end{equation*}
or $\mathbb P_j (d_{k+1}|\hat D_k) \to 0$ for $j=0,1$,
almost everywhere on $A$.

We will show that $A$ has probability $0$.
Suppose that there exists $\omega\in A$ such that
\begin{equation*}
\lim_{k\to \infty} \frac{\mathbb P_1(d_{k+1}=d_{k+1}(\omega)|\hat D_{k}=\hat D_{k}(\omega))}{\mathbb P_0(d_{k+1}=d_{k+1}(\omega)|\hat D_{k}=\hat D_{k}(\omega))}=1.
\end{equation*}
Note that $d_{k+1}(\omega)=0 \text{ or } 1$. Without loss of generality, consider the case where $d_{k+1}(\omega)=0$, we have
\begin{equation}
  \label{eq:t111}
\lim_{k\to \infty} \frac{\mathbb P_1(d_{k+1}=0|\hat D_{k}=\hat D_{k}(\omega))}{\mathbb P_0(d_{k+1}=0|\hat D_{k}=\hat D_{k}(\omega))}=1.
\end{equation}
Note that the statement $d_{k+1}=0$ is equivalent to
\begin{equation*}
   L_X(X_{k+1})L_{k}(\hat D_k)\leq \frac{\pi_0}{\pi_1}.
\end{equation*}
Because of the independence between $X_{k+1}$ and $\hat D_k$, we obtain 
\begin{align*}
&\mathbb P_j(d_{k+1}=0| \hat D_k=\hat D_k(\omega))=\\
& \mathbb P_j\left.\left(L_X(X_{k+1})L_k(\hat D_k)\leq \frac{\pi_0}{\pi_1}\right| \hat D_k=\hat D_k(\omega)\right)=\\
&\mathbb P_j\left(L_X(X_{k+1})L_k(\hat D_k(\omega))\leq \frac{\pi_0}{\pi_1} \right).
\end{align*}
Thus \eqref{eq:t111} is equivalent to
\begin{equation}
  \label{eq:12}
\lim_{k\to \infty} \frac{\mathbb P_1(L_X(X_{k+1})L_k(\hat D_k(\omega))\leq \frac{\pi_0}{\pi_1})}{\mathbb P_0(L_X(X_{k+1})L_k(\hat D_k(\omega))\leq \frac{\pi_0}{\pi_1})}=1.
\end{equation}
By \eqref{eq:mu} and the definitions of $\mathbb G_1$ and $\mathbb G_0$, \eqref{eq:12} is equivalent to 
\[
\lim_{k\to\infty} \frac{\mathbb G_1((1+L_k(\hat D_k(\omega)))^{-1})}{\mathbb G_0((1+L_k(\hat D_k(\omega))^{-1}) }=1.
\]
Because $\mathbb G_1$ and $\mathbb G_0$ are right-continuous, we have $\mathbb G_1/\mathbb G_0$ is also right-continuous. Moreover, $\mathbb G_1/\mathbb G_0$ is monotone non-decreasing. Therefore, we have
\[
\frac{\mathbb G_1((1+L_{\infty}(\omega))^{-1})}{\mathbb G_0((1+L_{\infty}(\omega))^{-1}) }=1.
\]
 However, this contradicts the dominance property (described earlier). We can use a similar argument to show that there does not exist $\omega$ such that $\mathbb P_j(d_{k+1}=d_{k+1}(\omega)|\hat D_k=\hat D_k(\omega))\to 0.$ Therefore, no such $\omega$ exists and this implies that $\mathbb P_0(A)=0$. Hence, $\mathbb P_0(L_{\infty}=0)=1$.
\end{IEEEproof}

\begin{Theorem}
Suppose that the flipping probabilities are bounded away from $1/2$. Then, $\mathbb P_e^k\to0$ as $k\to\infty$.
\end{Theorem}
\begin{IEEEproof}
We know that the likelihood ratio test states that $a_k$ decides $1$ if and only if $\bar b_k > 1-b_{k-1}$. The probability of deciding $1$ given that $H_0$ is true (Type~I error) is given by 
\begin{align*}
\mathbb P_{0}(d_k=1)&=\mathbb P_{0}(\bar b_k > 1-b_{k-1})\\
&=\mathbb E_0(1-\mathbb G_0(1-b_{k-1})).
\end{align*}
Since $L_{\infty}= 0$ almost surely, we have $b_k\to 0$ almost surely. We have \[
\lim_{k\to\infty} \mathbb P_{0}(d_k=1) =\lim_{k\to\infty}\mathbb E_0(1-\mathbb G_0(1-b_{k-1})).\]  By the bounded convergence theorem,  we have 
\begin{align*}
\lim_{k\to\infty} \mathbb P_{0}(d_k=1) &=1-\mathbb E_0(\lim_{k\to\infty} \mathbb G_0(1-b_{k-1}))\\
&= 1-\mathbb G_0(1)=0.
\end{align*}

Similarly, we can prove that $\lim_{k\to\infty} \mathbb P_1(d_k=0)=0$ (i.e., Type~II error probability converges to 0). Therefore, the error probability converges to 0.
\end{IEEEproof}

\begin{rem}[Additive Gaussian noise] Note that our convergence proof easily generalizes to the additive Gaussian noise scenario: Suppose that after $a_k$ makes a decision $d_k\in\{0,1\}$, it broadcasts a message $\hat d_k=F_k d_k +\mathcal N_k$ to other nodes, where $F_k\in(0,1)$ denotes a fading coefficient and $\mathcal N_k$ denotes zero-mean Gaussian noise.  Then, we can show that the error probability converges to 0 if $F_k$ are bounded away from 0 and the noise variances are bounded for all $k$. In other words, the signal-to-noise ratios are bounded away from 0.
\end{rem}

%
%
%

Now let us consider the convergence rate of the error probability. Without loss of generality, we assume that the prior probabilities are equal; i.e., $\pi_0=\pi_1=1/2$. The following analysis easily generalizes to unequal prior probabilities. Recall that $b_k=\mathbb P(\theta=H_1|\hat D_k)$ denotes the public belief.
It is easy to see that the error probability converges to 0 if and only if $b_k\to 0$ almost surely given $H_0$ is true and $ b_k\to 1$ almost surely given $H_1$ is true. 
Recall the proportionality property:
\[
\frac{d \mathbb G_1}{d \mathbb G_0} (r)=\frac{r}{1-r}.
\]
Moreover, we assume $\mathbb G_1$ and $\mathbb G_0$ are continuous and therefore under each of $H_0$ and $H_1$, the density of the private belief exists. By the above property, we can write these densities as follows:
\[f^1(r)=\frac{d \mathbb G_1}{d r}(r)=r  \rho(r),\] and \[f^0(r)=\frac{d \mathbb G_0}{d r}(r)=(1-r) \rho(r),\]
where $ \rho(r)$ is a non-negative function.

Without loss of generality, we assume that $H_0$ is the true hypothesis. Moreover, we assume that $\rho(1)>0$ and $\rho$ is continuous near $r=1$. This characterizes the behavior of the tail densities. We will generalize our analysis to polynomial tail densities later, where $\rho(r)\to 0$ as $r \to 1$.

The Bayesian update of the public belief when $\hat d_{k+1}=0$ is given by:
\begin{align}
\nonumber
&b_{k+1}=\mathbb P(\theta=H_1|\hat D_{k+1})\\
\nonumber
&=\frac{\mathbb P_1(\hat d_{k+1}=0|\hat D_k)b_k}{\sum_{j=0,1}\mathbb P_j(\hat d_{k+1}=0|\hat D_k)\mathbb P(\theta=H_j|\hat D_k)}\\
&=\frac{(q_k+(1-2q_k)\mathbb P_1(d_{k+1}=0|\hat D_k)) b_k}{\sum_{j=0,1} (q_k+(1-2q_k)\mathbb P_j(d_{k+1}=0|\hat D_k))\mathbb P(H_j|\hat D_k)}.
\label{eq:21}
\end{align}
It is easy to show that the public belief converges to 0 in the fastest rate if $\hat d_{k}=0$ for all $k$. We will establish the rate in this special case to bound the converge rate of the error probability.
Notice that $\mathbb P(\theta=H_1|\hat D_k)=b_k$ and $\mathbb P(\theta=H_0|\hat D_k)=1-b_k$. By Lemma 2, we have $L_k(\hat D_k)\to 0$ almost surely, under $H_0$. This implies that $b_k \to 0$ almost surely. If $b_k$ is sufficiently small, then we have
\begin{align}
\nonumber
\mathbb P_1(d_{k+1}=0|\hat D_k) &=1-\int_{1- b_k}^1 f^1(x)dx \\
& \simeq 1- \rho(1)( b_k-\frac{ b_k^2}{2})
\label{eq:p1}
\end{align}
and
\begin{align}
\nonumber
\mathbb P_0(d_{k+1}=0|\hat D_k) &=1-\int_{1- b_k}^1 f^0(x)dx \\
&\simeq 1- \rho(1)\frac{ b_k^2}{2}.\label{eq:p2}
\end{align}
Note that $\simeq$ means asymptotically equal.
We can also calculate the (conditional) Type~I error probability:
\begin{align}
\nonumber
\mathbb P_0(d_{k+1}=1|\hat D_k)&=1-\mathbb P_0(d_{k+1}=1|\hat D_k) \\
\nonumber
&=\int_{1- b_k}^1 f^0(x)dx\\
& \simeq  \rho(1)\frac{ b_k^2}{2}.
\label{eq:t1}
\end{align}
Note that \eqref{eq:t1} characterizes the relationship between the decay rate of Type~I error probability and the decay rate of $ b_k$.
Next we derive the decay rate of $b_k$.

Substituting \eqref{eq:p1} and \eqref{eq:p2} into \eqref{eq:21} and removing high order terms we obtain

\begin{align*}
 b_{k+1}\simeq \frac{(1-q_k) b_k-(1-2q_k) \rho(1) b_k^2}{(1-q_k)}.
\end{align*}
This implies that
\begin{align}
\label{eq:rec}
{ b_{k+1}}\simeq b_k\left(1-\frac{1-2q_k}{1-q_k}  \rho(1) b_k\right).
\end{align}
For any sequence that evolves according to \eqref{eq:rec}, the following lemma characterizes the convergence rate of the sequence.

\begin{lemma}
Suppose that a non-negative sequence $c_k$ satisfies $c_{k+1}=c_k(1-\delta c_k^n)$, where $n\geq 2$, $c_1<1$, and $\delta>0$. Then, for sufficiently large $k$, there exists two constants $C_1$ and $C_2$ such that
\[
\frac{C_1}{(\delta k)^{1/n}}\leq c_k \leq \frac{C_2}{(\delta k)^{1/n}}.
\] 
This implies that $c_k\to 0$ as $k\to\infty$ and $c_k=\Theta(k^{-1/n})$.
\end{lemma}
\begin{IEEEproof}
The proof is given in Appendix B. 
\end{IEEEproof}

\begin{Theorem}
Suppose that the flipping probabilities are bounded away from $1/2$ and $ \rho(1)$ is a non-negative constant. Then, the Type~I error probability converges to 0 as $\Omega(k^{-2})$.
\end{Theorem}

\begin{IEEEproof}
Using \eqref{eq:rec} and Lemma 3, we can get the convergence rate of the public belief conditioned on event that $\hat d_k=0$ for all $k$, in which case we have $b_k=\Theta(k^{-1})$. Recall that the public belief converges to 0 the fastest in this case among all possible outcomes. Therefore, we have $b_k=\Omega(k^{-1})$ almost surely.

Recall that $d_k=1$ if and only if $\bar b_k > 1-b_{k-1}$. Therefore, the Type I error probability is given by 
\begin{align}
\nonumber
\mathbb P_0(d_k=1)&=\mathbb P_0(\bar b_k > 1-b_{k-1}) \\
&=\mathbb E_0(1-\mathbb G_0(1-b_{k-1})). 
\label{eq:ss}\end{align}
Because $\rho$ is continuous at $1$, we have if $x<1$ is sufficiently close to 1, i.e., $1-x$ is positive and sufficiently small, then
\begin{align}
\nonumber
1-\mathbb G_0(x) &=\int_{x}^1 (1-x)\rho(x) dx  \\
\nonumber
& \geq \frac{\rho(1)}{2} \int_{x}^1 (1-x)dx \\
&=\frac{\rho(1)(1-x)^2}{4}.
\label{eq:sss}
\end{align}
From \eqref{eq:ss} and \eqref{eq:sss} and invoking Jensen's Inequality, we obtain
\begin{align}
\nonumber
\mathbb P_0(d_k=1) &\geq \frac{\rho(1)}{4}\mathbb E_0 [ b_{k-1}^2 ] \\
&\geq \frac{\rho(1)}{4}(\mathbb E_0 [b_{k-1}])^2. 
\label{eq:ssss}
\end{align}
Because $b_k=\Omega(k^{-1})$ almost surely, we have $\mathbb P_0(d_k=1)=\Omega(k^{-2})$.
\end{IEEEproof}

 Assume that $\rho(0)>0$ and $\rho$ is continuous at 0. Then, we can use the same method to calculate the decay rate of the Type~II error probability, which is the same as that of the Type~I error probability.
Note that the decay rate of the error probability depends linearly on $(1-2q_k)^{-2}$.

\subsubsection{Asymptotically uninformative nodes}
In this part, we consider the case where $q_k \to 1/2$ as $k \to \infty$, which means that the broadcasted decisions become asymptotically uninformative. Let \[Q_k= \frac{1-2q_k}{1-q_k}.\] Note that $q_k\to 1/2$ implies that $Q_k \to 0$. This parameter measures how ``informative'' the corrupted decision is: For example, if $q_k=0$ (where there is no flipping), then the decision is maximally informative in terms of updating the public belief. However if $q_k=1/2$, in which case $Q_k=0$, then the decision is completely uninformative in terms of updating the public belief.

We will derive a necessary condition on the decay rate of $Q_k$ to $0$ for the public belief $b_k$ to converge to $0$ under $H_0$, which gives us a necessary condition on $Q_k$ for asymptotic learning. For any sequence that evolve according to \eqref{eq:rec}, the following lemma characterizes necessary and sufficient conditions such that the sequence converges to 0.

\begin{lemma}Suppose that a non-negative sequence $\{c_k\}$ follows $c_{k+1}=c_k(1-\delta_k c_k^n)$, where $n\geq 1$, $c_1>0$, and $\delta_k> 0$. Then, $c_k$ converges to 0 if and only if there exists $k_0$ such that $\sum_{k=k_0}^{\infty}\delta_k=\infty$.
\end{lemma}
\begin{IEEEproof}
We will use the following claim to prove the lemma: For a non-negative sequence satisfying $c_{k+1}=c_k(1-r_k),$ where $c_1> 0$ and $r_k\in [0,1)$, we have $c_k \to 0$ if and only if there exists $k_0$ such that $\sum_{k=k_0}^{\infty} r_k=\infty$. To show this claim, we have \[c_{k+1}=c_1 \prod_{i=1}^k (1-r_i).\] Applying natural logarithm, we obtain\[
\ln c_{k+1}=\ln c_1 +\sum_{i=1}^k  \ln (1-r_i).
\]
From the above equation, we have $c_k\to 0$ if and only if $\sum_{i=1}^\infty  \ln (1-r_i)=-\infty$. In the case where there exists a subsequence of $\{r_k\}$ such that the subsequence is bounded away from 0, we have $\sum_{i=1}^{\infty}  \ln (1-r_i)=-\infty$. Therefore, $c_k\to 0$ as $k\to\infty$.
In the case where $r_k\to 0$, there exists $k_0$ such that $r_i\leq-\ln (1-r_i) \leq 2 r_i$ for all $i\geq k_0$. Therefore, we have $c_k\to 0$ if and only if $\sum_{k=k_0}^{\infty} r_k=\infty.$

We now show the lemma. First we show that the condition is necessary.  Suppose that $c_k\to 0$. Then, we have $\sum_{k=1}^{\infty} \delta_k c_k^n=\infty$. Since $c_k <1$, we have $\sum_{k=1}^{\infty} \delta_k =\infty$. Second we show by contradiction that the condition is sufficient. Suppose that there exist $k_0$ such that $\sum_{k=k_0}^{\infty} \delta_k =\infty$ and $c_k$ does not converge to 0. Since $c_k$ is monotone decreasing, $c_k$ must converge to a nonzero limit $c$. Therefore, for all $k$, we have $c_k \geq c$. Then, we have $c_{k+1}\leq c_k (1-\delta_k c^n)$. We have \[\sum_{k=k_0}^{\infty} \delta_k c^n =c^n \sum_{k=k_0}^{\infty} \delta_k =\infty.\] Therefore, we have $c_k\to 0$.
\end{IEEEproof}


\begin{Theorem}
Suppose that there exists $p>1$ such that \[Q_k =O\left(\frac{1}{k(\log k)^p}\right).\]
Then, the public belief converges to a nonzero limit almost surely.

\end{Theorem}
\begin{IEEEproof}
Suppose that there exists $p>1$ such that $Q_k =O\left({1}/({k(\log k)^p)}\right).$
Then, we have \[\sum_{k=2}^{\infty} Q_k<\infty.\] Therefore, by Lemma~4, $b_k$ in \eqref{eq:rec} does not converge to 0. Recall that \eqref{eq:rec} represents the recursion of $b_k$ conditioned on the event that the node broadcast decisions are all 0. Therefore, the public belief is the smallest among all possible outcomes. Hence, the public belief converges to a nonzero limit almost surely. 
\end{IEEEproof}

By \eqref{eq:ssss}, it is evident that if $b_k$ converges to a nonzero limit almost surely, then $\mathbb P_0(d_k=1)$ is bounded away from 0 and $\mathbb P_0(d_k=0)$ is bounded away from 1. Therefore, the system does not asymptotically learn the underlying truth. Hence Theorem 7 provides a necessary condition for asymptotically learning.

Theorem~7 also implies that for there to be a nonzero probability that the public belief converges to zero, we must have that there exists $p\leq 1$ such that $Q_k=\Omega(1/k(\log k)^p)$. If the public belief does not converge to zero, then it is impossible for there to be an eventual collective arrival at the true hypothesis. To explain this further, 
Let $\mathcal H$ denote the event that there exists a (random) $k_0$ such that the sequence of decisions $d_k=0$ for all $k\geq k_0$. 
Occurrence of this event signifies that after a finite number of decisions, the agents arrive at the true underlying state. Such an outcome also means that, eventually, each agent's private signal is overpowered by the past collective true verdict, so that a false decision is never again declared. In the literature on social learning, this phenomenon is called \emph{information cascade}  (e.g., \cite{WC}) or \emph{herding} (e.g., \cite{Smith}).  We use $\mathcal L$ to denote the event $\{b_k\to 0\}$.
Notice that $\mathcal H$ occurs only if $\mathcal L$ occurs.  Hence, $\mathcal H$ is a subset of the event that $b_k\to 0$, i.e., $\mathcal H \subset \mathcal L$.   These leads to the following corollary of Theorem~7.

\begin{Cor}
If $Q_k=O(1/k(\log k)^p)$ for some $p>1$, then $\mathbb P(\mathcal H)=0$.
\end{Cor}

So, by the corollary above, only if  $Q_k=\Omega(1/k(\log k)^p)$ for some $p\leq 1$ can we hope for there to be a nonzero probability that $b_k\to 0$ and thus of information cascade to the truth. Even under the situation that $b_k\to 0$, i.e., conditioned on $\mathcal L$, we expect that the \emph{rate} at which $b_k\to 0$ depends on the scaling law of $Q_k$.  The following theorem relates the scaling laws of $\{Q_k\}$ with those of $\{b_k\}$ and the Type I error probability sequence $\{\mathbb P_0(d_k=1)\}$. 
\begin{Theorem} Conditioned on $\mathcal L$, we have the following:
\begin{itemize}
\item[(i)]
Suppose that $Q_k=\Theta(1/k^{1-p})$ where $p \in (0,1)$. Then, $ b_k=\Omega(k^{-p})$ almost surely and $\mathbb P_0(d_k=1) =\Omega(k^{-2p})$. 
\item[(ii)] Suppose that $Q_k=\Theta(1/k)$. Then, $ b_k=\Omega(1/\log k)$ almost surely and $\mathbb P_0(d_k=1)=\Omega(1/(\log k)^2)$.

\item[(iii)] Suppose that $Q_k=\Theta\left({1}/({k(\log k)^p)}\right)$ where $p\in (0,1)$. Then, $b_k=\Omega(1/(\log k)^q)$ almost surely, where ${1}/{q}+{1}/{p}=1$, and $\mathbb P_0(d_k=1)=\Omega(1/(\log k)^{2q})$. 
\item[(iv)] Suppose that  $Q_k=\Theta\left({1}/({k\log k})\right)$. Then, $b_k=\Omega(1/\log\log k)$ almost surely and $\mathbb P_0(d_k=1) =\Omega(1/(\log \log k)^{2})$.

\end{itemize}
\end{Theorem}
\begin{IEEEproof}
The proof is given in Appendix C.
\end{IEEEproof}

Note that Theorem 8 provides upper bounds for the convergence rates of the public belief and error probability. However, recall that $\mathcal H$ is a subset of the event that $b_k\to 0$. Therefore, even if $b_k \to 0$ with certain probability, the probability of $\mathcal H$ is not guaranteed to be nonzero. Next we provide a necessary condition such that the probability of $\mathcal H$ is nonzero.

\begin{Theorem}
Suppose that there exists $p\leq 1$ such that \[Q_k =O\left(\frac{(p+\log k)(\log k)^{p-1}}{ (k( \log k)^{p})^{1/2}}\right).\]
Then, we have $\mathbb P(\mathcal H)=0$.

\end{Theorem}
\begin{IEEEproof}
We first state a key lemma which is a corollary of the Borel-Cantelli lemma \cite{pro}.
Consider a probability space $(S, \mathcal S, \mathcal P)$ and a sequence of events $\{\mathcal E_k\}$ in $\mathcal S$. We define the limit superior of $\{\mathcal E_k\}$ as follows:
\[
\limsup_{k\to \infty} \mathcal E_k \equiv \bigcap_{k=1}^{\infty} \bigl( \bigcup_{n= k} \mathcal E_n\bigr).
\]
Note that this is the event that infinitely many of the $\mathcal E_k$ occur. We use $\mathcal E_k^C$ to denote the complement of $\mathcal E_k$.
\begin{lemma} Suppose that \[\sum_{k=1}^{\infty}\mathcal P(\mathcal E_k|\mathcal E_{k-1}^C,\mathcal E_{k-2}^C,\ldots,\mathcal E_{1}^C)=\infty.\]
Then, \[\mathcal P(\limsup_{k\to \infty} \mathcal E_k )=1.\]

\end{lemma}

The proof of this lemma is omitted. Now we prove the theorem. Let $\mathcal E_k$ be the event that $d_k=1$, i.e., $a_k$ makes the wrong decision given $H_0$. Notice that $\mathcal E_k^C$ is the event that $d_k=0$. If \[Q_k =O\left(\frac{(p+\log k)(\log k)^{p-1}}{ (k( \log k)^{p})^{1/2}}\right),\] then using the similar analysis as those in Theorem 8, we have 
\[
\mathbb P_0(\mathcal E_k |\mathcal E_{k-1}^C,\mathcal E_{k-2}^C,\ldots,\mathcal E_{1}^C)=\Omega\left(\frac{1}{k(\log k)^p}\right).
\]
This implies that these terms are not summable, i.e., $\sum_{k=1}^{\infty} \mathbb P_0(\mathcal E_k |\mathcal E_{k-1}^C,\mathcal E_{k-2}^C,\ldots,\mathcal E_{1}^C)=\infty$. Therefore we have $\mathbb P_0(\limsup_{k\to \infty} \mathcal E_k)=1$, which means that with probability 1, $d_k=1$ occurs for infinitely many $k$. Consequentially, we have $\mathbb P_0(\mathcal H)=0$. By symmetry, $\mathbb P_1(\mathcal H)=0$. This concludes the proof.
\end{IEEEproof}

Suppose that the flipping probability converges to $1/2$ sufficiently fast. Then, even if the public belief converges to 0, its convergence rate is very small because the broadcasted decisions become uninformative in a fast rate. In this case, the private signals are capable to overcome the public belief infinitely often because of the slow convergence rate of the public belief.

\subsubsection{Polynomial tail density}
We now consider the case where the private belief has polynomial tail densities, that is, $\rho(r) \to 0$ as $r\to 1$ and there exist constants $\beta,\gamma>0$ such that
\begin{align}
\label{eq:tail}
\lim_{ r \to 1} \frac{ \rho(r)}{(1- r)^{\beta}} =\gamma.
\end{align}
Note that $\beta$ denotes the leading exponent of the Taylor expansion of the density at 1. The larger the value of $\beta$, the thiner the tail density. Note that Theorem 7 (necessary condition for $\mathbb P(\mathcal L) >0$) which was stated under the constant density assumption is also valid in the polynomial tail density case. 
 We can use the similar analysis as before to derive the explicit relationship between the convergence rate of $Q_k$ and the convergence rate of the public belief conditioned on $\mathcal L$. 
The following theorem establishes the scaling laws of the public belief and Type~I error probability for both uniformly informative and asymptotic uninformative cases. 

\begin{Theorem} 
Consider the polynomial tail density defined in \eqref{eq:tail}. 

\begin{itemize}
\item[1)] Uniformly informative case: Suppose that the flipping probabilities are bounded away from $1/2$. Then, we have $ b_k=\Omega(k^{-1/(\beta+1)})$ almost surely and $\mathbb P_0(d_k=1)=\Omega(k^{-(\beta+2)/(\beta+1)})$.
\item[2)] Asymptotically uninformative case: Suppose that the flipping probabilities converge to $1/2$, i.e., $Q_k \to 0$. Conditioned on $\mathcal L$, we have
 \begin{itemize}
\item[(i)] if $Q_k=\Theta(1/k^{1-p})$ where $p \in (0,1)$, then $ b_k=\Omega(k^{-p/(\beta+1)})$ almost surely and $\mathbb P_0(d_k=1)=\Omega(k^{-(\beta+2)p/(\beta+1)})$,
\item[(ii)] if $Q_k=\Theta(1/k)$, then $ b_k=\Omega((\log k)^{-1/(\beta+1)})$ almost surely and $\mathbb P_0(d_k=1)=\Omega((\log k)^{-(\beta+2)/(\beta+1)})$,
\item[(iii)] if $Q_k=\Theta\left({1}/({k(\log k)^p)}\right)$ where $p\in (0,1)$, then $b_k=\Omega((\log k)^{-q/(\beta+1)})$ almost surely, where ${1}/{q}+{1}/{p}=1$, and $\mathbb P_0(d_k=1)=\Omega((\log k)^{-(\beta+2)q/(\beta+1)})$,
\item[(iv)] if  $Q_k=\Theta\left({1}/({k\log k})\right)$, then $b_k=\Omega((\log\log k) ^{-1/(\beta+1)})$ almost surely and \\$\mathbb P_0(d_k=1)=\Omega((\log \log k)^{-(\beta+2)/(\beta+1)})$.
\end{itemize}
\end{itemize}

\end{Theorem}
\begin{IEEEproof}
The proof is given in Appendix D.
\end{IEEEproof}

Next we provide a necessary condition such that $\mathcal H$ has nonzero probability.
\begin{Theorem}
Suppose that there exists $p\leq 1$ such that \[Q_k =O\left(\frac{(p+\log k)(\log k)^{p-1}}{ (k(\log k)^p)^{1/(\beta+2)}}\right).\]
Then, we have $\mathbb P(\mathcal H)=0$.

\end{Theorem}
\begin{IEEEproof}
The proof is similar with that of Theorem 9 and is omitted.
\end{IEEEproof}

Note that as $\beta$ gets larger, this necessary condition states that $Q_k$ has to decay very slowly in order that it is possible for $\mathcal H$ to occur. 

Similarly we can calculate the decay rate for the Type~II error probability $\mathbb P_1(d_k=0)$. Assume that the tail density is given by \[\lim_{r\to0} \frac{\rho(r)}{r^{\bar \beta}} =\bar \gamma\] where $\bar \beta , \bar \gamma >0$. Then, we can show that if the flipping probabilities are bounded away from $1/2$, then \[\mathbb P_1(d_k=0)=\Omega(k^{-(\bar \beta+2)/(\bar \beta+1)}).\] The decay rate of the error probability is given by \[
\mathbb P_e^k=\Omega\left(k^{-(1+1/(\max{(\beta,\bar \beta)}+1))}\right).
\]

\section{Concluding Remarks}

We have studied the sequential hypothesis testing problem in two types of broadcast failures: erasure and flipping. In both cases, if the memory sizes are bounded, then there does not exist a decision strategy such that  the error probability converges to 0. In the case of random erasure, if the memory size goes to infinity, then there exists a decision strategy such that the error probability converges to 0, even if the erasure probability converges to 1. We also characterize explicitly the relationship between the convergence rate of the error probability and the convergence rate of the memory. In the case of random flipping, if each node observes all the previous decisions, then with the myopic decision strategy, the error probability converges to 0, when the flipping probabilities are bounded away from $1/2$. In the case where the flipping probability converges to $1/2$, we derive a necessary condition on the convergence rate of the flipping probability such that the error probability converges to 0. We also characterize explicitly the relationship between the convergence rate of the flipping probability and the convergence rate of the error probability. Finally, we have derived a necessary condition such that the event herding has nonzero probability.

Our analysis leads to several open questions. We expect that our results can be extended to multiple hypotheses testing problem, paralleling a similar extension in \cite{Kop}. In the case of random flipping, we have not studied the case where the memory size goes to infinity but each node cannot observe all the previous decisions. 
We also want to generalize the techniques used in this paper to more general network topologies. Moreover, besides erasure and flipping failures, we expect that our techniques can be used in the additive Gaussian noise scenario. With finite signal-to-noise ratios (SNR), the martingale convergence proof in Lemma 2 easily generalizes to this scenario. However, if SNR goes to 0 (e.g., the fading coefficient goes to 0, the noise variance goes to infinity, or the broadcasting signal power goes to 0), it is obvious that the convergence of error probability is not always true. We want to derive necessary and sufficient conditions on the convergence rate of SNR such that the error probability still converges to~0.

\appendices

\section{Proof of Theorem 3}
W extend the proof to the case where each node observes $m_k\geq 1$ previous decisions. The likelihood ratio test in this case is given by
\begin{equation*}
d_k=
\begin{cases} 1 & \text{ if } L_X(X_k)> t(\hat d_{k-1},\ldots, \hat d_{k-m_k}),
\\
0 & \text{ if } L_X(X_k)\leq t(\hat d_{k-1},\ldots, \hat d_{k-m_k}),
\end{cases}
\end{equation*}
where $ t(\hat d_{k-1},\ldots, \hat d_{k-m_k})=t_k/L_D^k(\hat d_{k-1},\ldots, \hat d_{k-m_k})$ denotes the testing threshold.
Among all possible combinations of $\{\hat d_{k-1},\ldots, \hat d_{k-m_k}\}$, it suffices to assume that the likelihood ratio in the case where each decision equals 0 (denoted by $\mathbf 0^{m_k}$) is the smallest and that in the case where each decision equals 1 (denoted by $\mathbf 1^{m_k}$) is the largest. Otherwise, we can always find the smallest and largest likelihood ratio. The case where the likelihood ratios for all possible combinations are equal can be excluded because it means the decisions observed have no useful information for hypothesis testing; and the node has to make a decision based on its own measurement, in which case the error probability does not converge to 0.

From these, we can define the Type~I and II error probabilities as in \eqref{eqex1} and \eqref{eqex2}.
\begin{figure*}[!t]
\normalsize
\begin{align}
\label{eqex1}
\nonumber
\mathbb P_0(d_k=1)&=\mathbb P_0(L_X(X_k)> t_k(\mathbf 0^m_k))\mathbb P_0(\hat d_{k-1}=0,\hat d_{k-2}=0,\ldots,\hat d_{k-m_k}=0) \\ 
\nonumber
&\quad+\mathbb P_0(L_X(X_k)> t_k(1,0,0,\ldots,0))\mathbb P_0(\hat d_{k-1}=1,\hat d_{k-2}=0,\ldots,\hat d_{k-m_k}=0) +\ldots \\
&\quad+\mathbb P_0(L_X(X_k)> t_k(\mathbf 1^{m_k}))\mathbb P_0(\hat d_{k-1}=1,\hat d_{k-2}=1,\ldots,\hat d_{k-m_k}=1) \\
\nonumber
&=\mathbb P_0(L_X(X_k)> t_k(\mathbf 0^{m_k})) +\mathbb P_0(t_k(1,0,0,\ldots,0)< L_X(X_k)\leq t_k(\mathbf 0^{m_k})) \\
\nonumber
 &\quad\mathbb P_0(\hat d_{k-1}=1,\hat d_{k-2}=0,\ldots,\hat d_{k-m_k}=0) +\ldots \\
 \nonumber
&\quad+\mathbb P_0(t_k(\mathbf 1^{m_k})< L_X(X_k)\leq t_k(\mathbf 0^{m_k}))\mathbb P_0(\hat d_{k-1}=1,\hat d_{k-2}=1,\ldots,\hat d_{k-m_k}=1)
\end{align}
and
\begin{align}
\label{eqex2}
\nonumber
\mathbb P_1(d_k=0)&=\mathbb P_1(L_X(X_k)\leq t_k(\mathbf 0^{m_k}))\mathbb P_1(\hat d_{k-1}=0,\hat d_{k-2}=0,\ldots,\hat d_{k-m_k}=0) \\ 
\nonumber
&\quad+\mathbb P_1(L_X(X_k)\leq t_k(1,0,0,\ldots,0))\mathbb P_1(\hat d_{k-1}=1,\hat d_{k-2}=0,\ldots,\hat d_{k-m_k}=0) +\ldots \\
\nonumber
&\quad+\mathbb P_1(L_X(X_k)\leq t_k(\mathbf 1^{m_k}))\mathbb P_1(\hat d_{k-1}=1,\hat d_{k-2}=1,\ldots,\hat d_{k-m_k}=1) \\
&=\mathbb P_1(t_k(\mathbf 1^{m_k})< L_X(X_k)\leq t_k(\mathbf 0^{m_k}))\mathbb P_1(\hat d_{k-1}=0,\hat d_{k-2}=0,\ldots,\hat d_{k-m_k}=0) \\
\nonumber
&\quad+\mathbb P_1(t_k(\mathbf 1^{m_k})< L_X(X_k)\leq t_k(1,0,0,\ldots,0)) \mathbb P_0(\hat d_{k-1}=1,\hat d_{k-2}=0,\ldots,\hat d_{k-m}=0) +\ldots \\
\nonumber
&\quad+\mathbb P_1( L_X(X_k)\leq t_k(\mathbf 1^{m_k})).
\end{align}

\hrulefill
\vspace*{4pt}
\end{figure*}

With the similar argument as that in the tandem network case, we have
$
\mathbb P_e^k=\pi_0 \mathbb P_0(d_k=1) +\pi_1 \mathbb P_1(d_k=0).
$
Suppose that $\mathbb P_e^k\to 0$ as $k\to\infty$. Then, we must have $\mathbb P_0(L_X(X_k)> t_k(\mathbf 0^{m_k}))\to 0$ and $\mathbb P_1(L_X(X_k)\leq t_k(\mathbf 1^{m_k})) \to 0$. Recall that $\mathbb P_0^X$ and $\mathbb P_1^X$ are equivalent measures. Hence we have $\mathbb P_j(t_k(\mathbf 1^{m_k}) < L_X(X_k)\leq t_k(\mathbf 0^{m_k}))\to 1$ for $j=0,1$. We have
 \begin{align*}&\mathbb P_j(\hat d_{k-1}=j_{k-1}, \hat d_{k-2}=j_{k-2},\ldots,\hat d_{k-m_k}=j_{k-m_k})=\\
 \vspace{-2cm}&\mathbb P_j(\hat d_{k-1}=j_{k-1}|\hat d_{k-2}=j_{k-2},\ldots, \hat d_{k-m_k}=j_{k-m_k})\cdot\\
 &\mathbb P_j(\hat d_{k-2}=j_{k-2}|\hat d_{k-3}=j_{k-3},\ldots, \hat d_{k-m_k}=j_{k-m_k})\cdot\\
 &\ldots \mathbb P_j(\hat d_{k-m_k+1}=j_{k-m_k+1}|\hat d_{k-m_k}=j_{k-m_k})\cdot \\
 &\mathbb P_j(\hat d_{k-m_k}=j_{k-m_k}).\end{align*}
We already know that $\mathbb P_j(\hat d_{k-m_k}=j_{k-m_k})$ is bounded away from 0 by $q_k$. Similarly, we can show
\begin{align*}
&\mathbb P_j(\hat d_{k-i}=j_{k-i}|\hat d_{k-i-1}=j_{k-i-1},\ldots, \hat d_{k-m_k}=j_{k-m_k}) \\
&=(1-q_k)\mathbb P_j( d_{k-i}=j_{k-i}|\ldots, \hat d_{k-m_k}=j_{k-m_k})\\
&\quad+q_k(1-\mathbb P_j( d_{k-i}=j_{k-i}|\ldots, \hat d_{k-m_k}=j_{k-m_k}))\\
&=q_k+(1-2q_k)\mathbb P_j( d_{k-i}=j_{k-i}|\ldots, \hat d_{k-m_k}=j_{k-m_k}).
\end{align*}
Hence $\mathbb P_e^k$ is also bounded below by $q_k^{m_k} \geq q_k^C$. This contradiction implies that $\mathbb P_e^k$ does not converge to 0 with any decision strategy.

\vspace{-0.2in}

\section{Proof of Lemma 3}
First it is easy to see that $c_k\to0$ because it is the only fixed point of the recursion. To show the convergence rate, we treat the recursion \eqref{eq:rec} as an ordinary difference equation (ODE). Therefore, we have
\[
\frac{d c_k}{d k} =-\delta c_k^{n+1}.
\]
The solution to this ODE is for some $C>0$
\[
c_k=\frac{C}{(\delta k)^{1/n}}.
\]

Therefore, for sufficiently large $k$, there exists two constants $C_1$ and $C_2$ such that
\[
\frac{C_1}{(\delta k)^{1/n}}\leq c_k \leq \frac{C_2}{(\delta k)^{1/n}}.
\]
which implies that 
\[
c_k=\Theta(k^{-1/n}).
\]

\section{Proof of Theorem 8}
(i). Suppose that $Q_k=\Theta(1/k^{1-p})$ where $p \in (0,1)$. Conditioned on $\mathcal H$, we have recursion \eqref{eq:rec} for the public belief $b_k$. Using this recursion, we can get similar results as those in Lemma 3, that is, there exists $C_1>0$ and $C_2>0$ such that 
\begin{align}
\frac{C_1}{k Q_k} \leq  b_k \leq \frac{C_2}{k Q_k}.
\label{eq:ap}
\end{align}
Plugging in the convergence rate of $Q_k$ in \eqref{eq:ap} establishes the claim.

(ii)-(iv). Suppose that $Q_k=\Theta(1/k (\log k)^p)$, where $p \in [0,1]$. Then, by \eqref{eq:rec}, we have
\[
b_{k+1}-b_k = \frac{C b_k^2}{k(\log k)^p}
\]
for some constant $C>0$.
For $p=0$, the solution to this ODE satisfies $b_k=\Theta(1/\log k)$, which proves (ii). When $p\in (0,1)$, the solution satisfies $b_k=\Theta(1/(\log k)^q)$, where $1/q+1/p=1$. This establishes (iii). Finally, when $p=1$, the solution satisfies $b_k=\Theta(1/\log\log k)$. Note that all these rates are derived conditioned on $\mathcal H$. By the fact that conditioned on $\mathcal H$, the decay rate is the fastest among all outcomes, we obtain the desired results.
Having established the convergence rate of $b_k$, the convergence rate for the error probability in each claim follows from~\eqref{eq:ssss}.

\section{Proof of Theorem 10}
Proof of claim 1: If the flipping probabilities are bounded away from $1/2$, then the public belief $b_k$ converges to 0 and conditioned on $\mathcal H$ we have
\begin{align}
\nonumber
\mathbb P_1(d_{k+1}=0|\hat D_k) &=1-\int_{1- b_k}^1 f^1(x)dx \\
&\simeq 1- \frac{\gamma}{\beta} b_k^{\beta+1}
\label{eq:p13}
\end{align}
and
\begin{align}
\nonumber
\mathbb P_0(d_{k+1}=0|\hat D_k) &=1-\int_{1- b_k}^1 f^0(x)dx \\
&\simeq 1- \frac{\gamma}{\beta+1}b_k^{\beta+2}.\label{eq:p23}
\end{align}
We can also calculate the (conditional) Type~I error probability in this case:
\begin{align}
\nonumber
\mathbb P_0(d_{k+1}=1|\hat D_k)&=1-\mathbb P_0(d_{k+1}=1|\hat D_k) \\
\nonumber
&=\int_{1- b_k}^1 f^0(x)dx\\
& \simeq  \frac{\gamma}{\beta+1}b_k^{\beta+2}.
\label{eq:t11}
\end{align}
Note that \eqref{eq:t11} describes the relationship between the decay rate of Type~I error probability and the decay rate of $ b_k$.
Next we derive the decay rate of $b_k$.

By \eqref{eq:p13} and \eqref{eq:p23}, we can derive the recursion for the public belief as follows:
\begin{align}
 b_{k+1}= b_k-\frac{\gamma}{\beta}Q_k  b_k^{\beta+2}.
\label{eq:reccc}
\end{align}
By Lemma 3, we know that $ b_k\to 0$ and the decay rate is $b_k=\Theta(k^{-{1}/{(\beta+1)}}).$
Recall that conditioned on the event that $\hat d_k=0$ for all $k$, the convergence of $b_k$ is the fastest. Therefore, we have $b_k=\Omega(k^{-{1}/{(\beta+1)}})$ almost surely.
From \eqref{eq:t11} and invoking Jensen's Inequality, we obtain
\begin{align}
\nonumber
\mathbb P_0(d_k=1)&\geq \frac{\gamma}{\beta+1}\mathbb E_0 [ b_k^{\beta+2} ] \\
&\geq \frac{\gamma}{\beta+1}(\mathbb E_0 [b_k])^{\beta+2}. 
\end{align}
Because $b_k=\Omega(k^{-1/(\beta+1)})$ almost surely, we have $
\mathbb P_0(d_k=H_1)= \Omega(k^{-{(\beta+2)}/{(\beta+1)}}).
$

Proof of claim 2: Using Lemma 3, we can show that there exist two positive constants $C_1$ and $C_2$ such that
\begin{align}
\frac{C_1}{(kQ_k)^{1/(\beta+1)}}\leq  b_k \leq \frac{C_2}{(kQ_k)^{1/(\beta+1)}}.
\label{eq:c}
\end{align}
Therefore, if $Q_k = 1/k^{1-p}$, then using \eqref{eq:c} and the fact that $b_k$ given $\mathcal H$ is the smallest among all possible outcomes, we have $b_k=\Omega(k^{-p/(\beta+1)})$. This establishes (i).
For (ii)-(iv), we can solve the ODEs given by \eqref{eq:reccc} and the solutions give rise to the convergence rates for $b_k$, which in turn characterize the convergence rates of the error probabilities.

\section*{Acknowledgment}
The authors wish to thank the anonymous reviewers for the careful reading of the manuscript and constructive comments that have improved the presentation.

\bibliographystyle{IEEEbib}

\end{document}